\documentclass[submission, Phys]{SciPost}
\usepackage{doi,hyperref}
\hypersetup{colorlinks = true}
\usepackage{amsmath,amsthm,amssymb}
\usepackage{color}
\definecolor{pink}{rgb}{0.85,0.01,0.78}

\newcommand{\Qone}{Q^{(1)}}
\newcommand{\Qtwo}{Q^{(2)}}

\usepackage[numbers,square,sort&compress]{natbib}
\newcommand{\so}{\scriptscriptstyle \rm I}
\newcommand{\st}{\scriptscriptstyle \rm I\hspace{-1pt}I}

\newcommand{\bx}{\bar x}

\newcommand{\by}{\bar y}
\newcommand{\bz}{\bar z}

\newcommand{\uc}{u^{\scriptscriptstyle C}}
\newcommand{\ub}{u^{\scriptscriptstyle B}}
\newcommand{\vc}{v^{\scriptscriptstyle C}}
\newcommand{\vb}{v^{\scriptscriptstyle B}}
\newcommand{\bu}{\bar u}
\newcommand{\bv}{\bar v}

\newcommand{\buc}{\bar{u}^{\scriptscriptstyle C}}
\newcommand{\bub}{\bar{u}^{\scriptscriptstyle B}}
\newcommand{\bvc}{\bar{v}^{\scriptscriptstyle C}}
\newcommand{\bvb}{\bar{v}^{\scriptscriptstyle B}}

\newcommand{\be}[1]{\begin{equation}\label{#1}}
\newcommand{\ba}[1]{\begin{multline}\label{#1}}
\newcommand{\ee}{\end{equation}}
\newcommand{\ea}{\end{eqnarray}}

\newcommand{\Res}{\mathop{\rm Res}}

\newtheorem{thm}{Theorem}[section]

\newtheorem{lemma}[thm]{Lemma}

\def\qed{\hfill\nobreak\hbox{$\square$}\par\medbreak}
\makeatletter
 \@addtoreset{equation}{section}
 \makeatother

\newcommand{\bea}{\begin{eqnarray}}
\newcommand{\eea}{\end{eqnarray}}

\begin{document}

\begin{center}
\begin{LARGE}
{\bf Master equation for correlation functions\\ in algebra symmetry $\mathfrak{gl}(2|1)$  related models }
\end{LARGE}

\vspace{8mm}

\begin{large}
A.~Hutsalyuk${}^{a}$, A.~Liashyk${}^{b,c}$\  \footnote{
hutsalyuk@gmail.com, a.liashyk@gmail.com}
\end{large}


${}^a$ {\it Department of Theoretical Physics, \\
  E\"otv\"os Lor\'and University Budapest
MTA-ELTE ``Momentum'' Integrable Quantum Dynamics Research Group,\\
  E\"otv\"os Lor\'and University Budapest}\\[1ex]

${}^b$ {\it National Research University Higher School of Economics, Faculty of Mathematics, Moscow, Russia}\\[1ex]

${}^c$ {\it Skolkovo Institute of Science and Technology, Moscow, Russia}\\[1ex]

\end{center}


\begin{abstract}

We consider integrable models solved by the nested algebraic Bethe ansatz and associated with $\mathfrak{gl}(2|1)$ or $\mathfrak{gl}(3)$ algebra symmetry. The analogue of sum formulae, previously formulated for scalar products, is established for the form factors and correlation functions. These formulae are direct generalisation of the some earlier results derived for models with $\mathfrak{gl}(2)$ symmetric $R$-matrix. It is also shown that in the case of algebra symmetry $\mathfrak{gl}(2|1)$ related models such formula allows to establish a multiple integral representation for correlation functions and form factors.

\end{abstract}


\subsection{Introduction}

Correlation functions and form factors of integrable systems were the object of interest for a long time \cite{Honerkamp, Jimbo1, Jimbo2, Boos1}. Among the different approaches one of the most successful is {\it algebraic Bethe ansatz} (ABA) developed in \cite{Faddeev2, Faddeev3, Faddeev4}. Using the ABA correlation functions of integrable systems were extensively studied \cite{Korepin1, Izergin1984, CTW2, KitMT00, KitanineMaiilletTerrasSlavnov2002, Q_series, Q_series_dynamical, GohKS04, KitanineMaiilletTerrasSlavnov2008}. Asymptotic behaviour, temperature and time dependence of correlation functions were established in multiple systems. 

Recently the major interest is attracted by models described by the {\it nested  Bethe ansatz} (see \cite{KulRes82, KulishSklyanin, Kulish85}). These models are related to multicomponent systems or systems with the additional internal degrees of freedom (such as spin, colour charge, etc.) \cite{Yang, YangSutherland, McGuire, TakahashiGas}. However up to now these systems were much less studied and there are relatively few results on correlation functions in case of models described by the nested ABA (see \cite{Kozlowski}, \cite{Pozsgay2020}). 

The problem is the complexity of the method. Thus, typically in order to build the $n$-point correlation function using the ABA the knowledge of at least the scalar product of two eigenvectors is required. Usually such  scalar products are given by extremely bulky expressions containing multiple summations (so-called Reshetikhin formula \cite{Res86,SLHRP2}) and it is quite complicated to find a compact form for them. In the case of one-component models (or the same, algebra symmetry $\mathfrak{gl}(2)$ related models) the problem was solved in \cite{Slavnov1}. In the case of multicomponent models (or the same, algebra symmetry $\mathfrak{gl}(N)$ related models) the problem remained unsolved for a long time. Recent advances in this direction \cite{SLHRP3, Liashyk} finally allow us to build an integral representation for correlation functions in algebra symmetry $\mathfrak{gl}(2|1)$ (graded algebra) related models. This is the main result of current work.

Note that in most of the paper we do not focus on a particular model, since algebraic Bethe ansatz approach allows one to describe simultaneously all models associated with a particular algebra symmetry. See section \ref{Bethe_ansatz} and the comment at the end.

The paper is organised as follows. In Section \ref{def_notation} we introduce notation and give a short description of the ABA. In section \ref{Reshetikhin}  we prove the sum formula for generation function of correlators, i.e. the analogue of the Reshetikhin formula for correlation functions. In section \ref{me_integral} integral representation for correlation functions in algebra symmetry $\mathfrak{gl}(2|1)$ related models are established. In section \ref{F_series} we briefly describe how the result of the previous section can be derived using the form factor series. In Conclusion the brief outlook of the perspectives is given.

\section{Definition and notation\label{def_notation}}

\subsection{Notation}
Through the paper the following functions are used
\be{shorthand_functions}
\begin{split}
&f(x,y)=\frac{x-y+c}{x-y},\qquad
g(x,y)=\frac{c}{x-y},\qquad 
h(x,y)=\frac{f(x,y)}{g(x,y)}=\frac{x-y+c}{c},\qquad\\
&t(x,y)=\frac{g(x,y)}{h(x,y)}=\frac{c^2}{(x-y)(x-y+c)}.
\end{split}
\ee
We use the shorthand notation for sets $\bu=\{u_1,\dots,u_k\}$.  Sets are marked with Arabic or Roman numerals or Greek letters, for example $\bu_1$, $\bv_{\st}$, $\bv_{\alpha}$, etc. Individual elements of the sets are labelled by Latin letters, for example $u_j$, $v_k$. We use the following notation for complements of the sets $\bu_j=\bu\setminus u_j$. For arbitrary functions $G(s)$, $F(s, t)$ and arbitrary sets $\bar x,\bar y$ the following notation are applied for the products
\be{shorthand_product}
\begin{split}
G(\bar y)=\prod_{j=1}^{\#\bar y}G(y_j),\qquad
F(s,\bar y)=\prod_{j=1}^{\#\bar y}F(s,y_j),\qquad F(\bar x,\bar y)=\prod_{j=1}^{\#\bar y}\prod_{k=1}^{\#\bar x}F(x_k,y_j),\qquad\mbox{etc.}
\end{split}
\ee
For $a=f,g,h$ the  short-hand notation for skew-symmetric products are used
\be{skew}
\Delta_a(\bar x)=\prod_{i>j}a(x_i,x_j),\qquad \Delta'_a(\bar x)=\prod_{i<j}a(x_i,x_j).
\ee

\subsection{Algebraic Bethe ansatz\label{Bethe_ansatz}}

In the method  of ABA the system is described by {\it monodromy matrix} $T$ that satisfies {\it RTT-relation} \cite{Faddeev2,  Faddeev3, Faddeev4} with a proper {\it $R$-matrix}
\be{RTT}
R_{12}(v,u) T_{01}(v)T_{02}(u)=T_{02}(u)T_{01}(v)R_{12}(v,u).
\ee
\eqref{RTT} holds in a tensor product of  {\it quantum space} 0 (coincides with a Hilbert space of the Hamiltonian of a system under consideration) and two {\it auxiliary spaces} $1$, $2$ that are $\mathbb C^3$ in the case of algebra symmetry $\mathfrak{gl}(3)$ and $\mathbb C^{2|1}$ in the case of algebra symmetry $\mathfrak{gl}(2|1)$ related models. Here $\mathbb C^{2|1}$ denotes $Z_2$-graded vector space with a grading  $[1]=[2]=0$, $[3]=1$ (square brackets denote the parity). Matrices acting in $\mathbb C^{2|1}$ in case of $\mathfrak{gl}(2|1)$ algebra symmetry are also graded with a grading given by $[e_{ij}]=[i]+[j]$ where we define elementary units $(e_{ij})_{ab}=\delta_{ai}\delta_{bj}$.  The $R$-matrix satisfies {\it Yang-Baxter equation} (YBE)
\be{YBE}
R_{12}(v,u) R_{13}(v)R_{23}(u)=R_{23}(u)R_{13}(v)R_{12}(v,u),
\ee
that holds in a tensor product of three spaces $\left(\mathbb C^{3}\right)^{\otimes 3}$ (or
$\left(\mathbb C^{2|1}\right)^{\otimes 3}$ in a case of the $\mathfrak{gl}(2|1)$ algebra symmetry). In this paper we consider the rational R-matrix
\be{Rmat}
R(u,v)=\mathbb I+\frac{c\mathbb P}{u-v},\qquad\qquad
\mathbb P=\sum_{i,j=1}^3(-1)^{[j]}e_{ij}\otimes e_{ji},\qquad \mathbb I=\sum_{i,j=1}^{3}e_{ii}\otimes e_{jj}.
\ee
(For algebra symmetry $\mathfrak{gl}(3)$ all $[j]=0$). We denote the Bethe ansatz vacuum by $|0\rangle$ and assume normalisation $\langle0|0\rangle=1$. Vacuum eigenvalues of the diagonal entries of the monodromy matrix\footnote{Here and further subscripts of $T_{ij}$ denote the matrix  indices in the auxiliary space, not  numbers of spaces.} $T_{ii}$ are denoted by $\lambda_i$ and their ratios by  $r_i$\footnote{Note that these are the only quantities that depend on a particular model in all the paper.}
\be{lambda}
T_{ii}(t)|0\rangle=\lambda_i(t)|0\rangle,\qquad i=1,\dots,3,
\ee
\be{rr}
r_1(t)=\lambda_1(t)/\lambda_2(t),\qquad r_3(t)=\lambda_3(t)/\lambda_2(t).
\ee
We expand agreement \eqref{shorthand_product}  for the commuting operators, for instance $T_{12}(\bu)$, $T_{22}(\bv)$, etc. Note, however, that in case of algebra symmetry $\mathfrak{gl}(2|1)$ related models operators $T_{23}$, $T_{32}$ with different arguments do not commute with themselves, thus $T_{23}(u)T_{23}(v)\ne T_{23}(v)T_{23}(u)$. Instead, in this case we introduce symmetrised products
\be{bT}
\mathbb T_{j3}(\bv)=\Delta_h(\bv)^{-1}T_{j3}(v_1)\dots T_{j3}(v_n),\qquad \mathbb T_{3j}(\bv)=\Delta_h'(\bv)^{-1}T_{3j}(v_1)\dots T_{3j}(v_n),\qquad j=1,2.
\ee

ABA implies the existence of the special objects called {\it Bethe vectors}. In a case of algebra symmetry $\mathfrak{gl}(2)$ related models Bethe vectors are monomials on the matrix element $T_{12}(u)$ acting on the vacuum and depend on a set of spectral parameters $\bu$ (also called Bethe parameters)
\be{BVgl2}
|\bu\rangle=T_{12}(u_a)\dots T_{12}(u_1)|0\rangle.
\ee
For models related to the higher rank algebra symmetries we need to apply the so-called {\it nested Bethe ansatz} procedure \cite{KulRes82, KulishSklyanin, Kulish85}. The Bethe vectors now are given by the special polynomials on monodromy matrix entries acting on vacuum, and depend on two sets of spectral parameters $\{\bu,\bv\}$.  We refer to these sets as the Bethe parameters of the first and the second level of the nesting. The explicit form of the Bethe vectors is given by \cite{SRP3}\footnote{Pay attention that our normalisation differs from used in \cite{SLHRP2,SLHRP3} by the additional factor $f(\bv,\bu)$ in the numerator.}
\be{BVgl21}
|\bu;\bv\rangle=\sum \frac{1}{\lambda_2(\bu)\lambda_2(\bv_{\st})}{g(\bv_{\so},\bu_{\so})f(\bu_{\so},\bu_{\st})g(\bv_{\st},\bv_{\so})h(\bu_{\so},\bu_{\so})}\mathbb T_{13}(\bu_{\so})T_{12}(\bu_{\st})\mathbb T_{23}(\bv_{\st})|0\rangle,
\ee
in the case of $\mathfrak{gl}(2|1)$ and \cite{SRP2}
\be{BVgl3}
|\bu;\bv\rangle=\sum \frac{1}{\lambda_2(\bv_{\st})\lambda_2(\bu)}K_n(\bv_{\so}|\bu_{\so})f(\bv_{\st},\bv_{\so})f(\bu_{\so},\bu_{\st})T_{13}(\bu_{\so})T_{12}(\bu_{\st})T_{23}(\bv_{\st})|0\rangle,
\ee
in the case of $\mathfrak{gl}(3)$ algebra symmetry  related models, where {\it Izergin-Korepin determinant} is defined as
\be{IK}
K_n(\bx|\by)=\Delta_g'(\bar x)\Delta_g(\bar y)h(\bar x,\bar y)\det_n\left[ t(x_j,y_k)\right],
\ee
and in both cases sum is taken over partitions $\bu\to\{\bu_{\so},\bu_{\st}\}$, $\bv\to\{\bv_{\so},\bv_{\st}\}$. Dual (left) Bethe vectors can be obtained by mapping  $\psi:\;|0\rangle\to\langle0|$, $\psi:\;T_{ij}\to (-1)^{[i][j]+[i]}T_{ji}$ and $\psi(AB)=(-1)^{[A][B]}\psi(B)\psi(A)$.

We denote the cardinalities of sets as $\#\bu=a$ and $\#\bv=b$. In the case sets $\{\bu,\bv\}$ satisfy the system of {\it Bethe ansatz equations} (BAE) Bethe vectors become eigenvectors of the Hamiltonian of the model. We call such Bethe vectors {\it on-shell}, otherwise they are called {\it off-shell} or generic. Using the shorthand notation Bethe equations can be written as
\be{BEgl3}
\begin{split}
&r_1(u_j)=\frac{1}{\varkappa_1}\frac{f(u_j,\bu_j)}{f(\bu_j,u_j)}f(\bv,u_j),\qquad j=1,\dots,a,\\
&r_3(v_j)=\frac{1}{\varkappa_2}\left(\frac{f(\bv_j,v_j)}{f(v_j,\bv_j)}\right)^sf(v_j,\bu),\qquad j=1,\dots,b,
\end{split}
\ee
where $\varkappa=\{\varkappa_1,\varkappa_2\}$ are {\it twists} (see \cite{Izergin1984,Korepin1}), $s=1$ for algebra symmetry $\mathfrak{gl}(3)$ and $s=0$ for algebra symmetry $\mathfrak{gl}(2|1)$ related models. In case when $\varkappa_1\ne1$ and/or $\varkappa_2\ne1$ BAE traditionally are explicitly called {\it twisted BAE} and on-shell Bethe vectors that correspond to solutions of such BAE are called {\it twisted on-shell-Bethe vectors}.

We define {\it (twisted) transfer matrix} as a (graded) trace of the monodromy matrix. 
\be{transfer}
t_{\varkappa}(w)=\sum_{i=1}^3\varkappa_i (-1)^{[i]}T_{ii}(w).
\ee
It posses the commutativity property for arbitrary parameters $u$, $v$
\be{t_commutation}
[t_{\varkappa}(u),t_{\varkappa}(v)]=0.
\ee
(Twisted) on-shell Bethe vectors are also eigenvectors of the (twisted) transfer matrix
\be{t_twisted}
t_{\varkappa}(w)|\bu;\bv\rangle=\tau_{\varkappa}(w|\bu;\bv)|\bu;\bv\rangle.
\ee

Further we use a concept of {\it two-site  model} \cite{Korepin1, Izergin1984} (also called  {\it partial model}). Consider the system of length $L$ with two subsystems, correspondingly $[0,x]$ and $[x,L]$.  We denote quantities belonging to the $i$ subsystem by the superscript $(k)$ and call them partial. Thus $\lambda_i^{(k)}(t)$, $i=1,\dots,3$, $k=1,2$ denote the vacuum eigenvalues of diagonal elements of the partial monodromy matrix belonging  to subsystem $k$ and the vacuum vector is given by $|0\rangle=|0\rangle^{(2)}\otimes|0\rangle^{(1)}$. It can be shown that Bethe vectors of the total model can be expressed via the partial Bethe vectors in a following way\footnote{The dual Bethe vector has absolutely similar property
\be{dual_partial_model}
\langle \bu;\bv|=\sum r_1^{(1)}(\bu_{\st})r_3^{(2)}(\bv_{\so})f(\bu_{\so},\bu_{\st})f(\bv_{\so},\bv_{\st})f(\bv_{\st},\bu_{\so})\langle\bu_{\st};\bv_{\st}|^{(2)}\otimes\langle\bu_{\so};\bv_{\so}|^{(1)}.
\ee}
\be{partial_model}
|\bu;\bv\rangle=\sum r_1^{(2)}(\bu_{\so})r_3^{(1)}(\bv_{\st})f(\bu_{\st},\bu_{\so})f(\bv_{\st},\bv_{\so})f(\bv_{\so},\bu_{\st})|\bu_{\st};\bv_{\st}\rangle^{(2)}\otimes|\bu_{\so};\bv_{\so}\rangle^{(1)}.
\ee
The sum is taken over partitions $\bu\to\{\bu_{\so},\bu_{\st}\}$, $\bv\to\{\bv_{\so},\bv_{\st}\}$.

Following \cite{Korepin1} we use the concept of {\it generalised model} in which sets $\{r_1(u_k)\}$,   $k=1,\dots,a$,  $\{r_3(v_j)\}$, $j=1,\dots,b$ are treated as sets of free parameters, without any reference to the particular model. 

As we already mention, we do not concentrate on a particular model, but rather on algebra symmetry of the $R$-matrix and our results in this way does not depend on particular model. In order to specify results for a particular model it is enough to specify $\lambda_i$ defined in \eqref{lambda}, that can be done at the very end.  Thus, solving the problem for $R$-matrix with algebra symmetry $\mathfrak{gl}(2|1)$ we can describe both 1D Fermi gas \cite{TakahashiGas,Yang} and lattice hopping model \cite{tJ,GohmannSeel}. Exception is section \ref{F_series} that concentrates on the particular vase of t-J model.

\subsection{Generation function}

Introduce operators $Q^{(k)}_i$, $k=1,2$, $i=1,2$ that count the numbers of particles of type $i$ in the $k$ subsystem (and $Q_i^{(1)}+Q_i^{(2)}=Q_i$). Operators 
\be{generator}
\exp(\alpha Q)=\exp\left(-\alpha_1Q_1^{(1)}+\alpha_2Q_2^{(1)}\right),\qquad \exp(\alpha Q^{(1)})=\exp\left(-\alpha_1Q_1^{(1)}+\alpha_2Q_2^{(2)}\right)
\ee
are generators for two-point correlation functions of the densities of particles. 

Thus, for the Gaudin-Yang model (spin-1/2 Fermi gas) with a Hamiltonian
\be{GY}
H=\int_0^L dx\sum_{\alpha,\beta}\left\{\partial\psi_{\alpha}^{\dagger}\partial\psi_{\alpha}+2c\psi_{\alpha}^{\dagger}\psi_{\beta}^{\dagger}\psi_{\beta}\psi_{\alpha}\right\},\qquad \alpha,\beta=\uparrow,\downarrow,
\ee
and canonical commutation relation $\{\psi^{\dagger}_{\alpha}(x),\psi(y)\}=\delta(x-y)\delta_{\alpha\beta}$ we have for the particles densities ($q_1$ is the total density and $q_2$ is the density of particles with the projection of spin down)
\be{Q_def}
\Qone_i=\int_{0}^x dz\; q_i(z),\qquad \Qtwo_i=\int_{x}^L dz\; q_i(z),
\ee
\be{Gen_func_gl_21}
\langle q_i(x)q_i(0)\rangle=-\frac{1}{2}\left.\frac{\partial^2}{\partial x^2}\frac{\partial^2}{\partial\alpha^2_i}\left\langle\exp\left(\alpha \Qone\right)\right\rangle\right|_{\alpha=0},
\ee
and for $i\ne j$
\be{Gen_func_gl_21_2}
\langle q_i(x)q_j(0)\rangle=\left.\frac{\partial^2}{\partial x^2}\frac{\partial^2}{\partial\alpha_1\partial\alpha_2}\left\langle\exp\left(-\alpha_1 \Qone_1+\alpha_2\Qtwo_2\right)\right\rangle\right|_{\alpha=0}.
\ee
Notation $\alpha=0$ here and further means that $\alpha_1=\alpha_2=0$. Since by definition $\Qtwo_2=Q_2-\Qone_2$ and $Q_2|\bub;\bvb\rangle=b|\bub;\bvb\rangle$, ($b$ is just a total number of particles on $[0,L]$) we can write 
\be{QtoQ1}
\exp(\Qone)=\left.\exp(\alpha Q)\right|_{\alpha_2\to-\alpha_2}\exp(\alpha_2Q_2). 
\ee
Thus we will concern only about one of generation functions, since the other one can be found via \eqref{QtoQ1}.

In the same way correlators in other models can be expressed via the derivatives of\\ $\langle \exp(\alpha \Qone)\rangle$, $\langle\exp(\alpha Q)\rangle$ (for example correlators of electrons densities in supersymmetric t-J model, densities in the Fermi-Bose mixtures). For the lattice models the derivatives w.r.t. $x$ would be naturally replaced by the finite differences and the integrals over the subsystems by the sums.

\section{Reshetikhin formula\label{Reshetikhin}}

We define scalar product of two Bethe vectors $S_{a,b}$ as
\be{scalp}
S_{a,b}\left(\buc;\bvc|\bub;\bvb\right)=\langle\buc;\bvc|\bub;\bvb\rangle.
\ee
It can be shown that scalar product of the off-shell Bethe vectors can be presented via {\it highest coefficients} (Reshetikhin formula)\footnote{We do not give here the explicit representation for $Z_{a,b}$ since we do not need them, but it can be found in \cite{SRP} for algebra symmetry $\mathfrak{gl}(3)$ related models and in \cite{SLHRP2} for algebra symmetry $\mathfrak{gl}(2|1)$ related models. For algebra symmetry $\mathfrak{gl}(2)$ related models they coincide with Izergin-Korepin determinant \eqref{IK}.} $Z_{k,n}$ \cite{Res86} as 
\begin{multline}
\label{scalp}
S_{a,b}\left(\buc;\bvc|\bub;\bvb\right)=\sum r_1(\bub_{\so})r_1(\buc_{\st})r_3(\bvb_{\so})r_3(\bvc_{\st})f(\buc_{\so},\buc_{\st})f(\bub_{\st},\bub_{\so})f(\bvc_{\st},\bvc_{\so})f(\bvb_{\so},\bvb_{\st})\\
\times f(\bvc_{\so},\buc_{\so})f(\bvb_{\st},\bub_{\st})Z_{a-k,n}(\buc_{\st};\bub_{\st}|\bvc_{\so};\bvb_{\so})Z_{k,b-n}(\bub_{\so};\buc_{\so}|\bvb_{\st};\bvc_{\st}).
\end{multline}
Here $\#\bu=a$, $\#\bv=b$. The sum is taken over partitions $\bub\to\{\bub_{\so},\bub_{\st}\}$, $\bvb\to\{\bvb_{\so},\bvb_{\st}\}$ and the same for $\{\buc,\bvc\}$. Coefficients $Z_{m,n}$ do not depend on a particular model but only on the algebra symmetry of $R$-matrix.

Our goal in this section is to show that for the matrix elements of operator $\exp(\alpha Q)$ formula with a structure similar to \eqref{scalp} can be established.

Using the property \eqref{partial_model} we immediately arrive at the following representation for $\langle\exp(\alpha Q)\rangle$
\begin{multline}
\label{coprod}
\langle \buc;\bvc |\exp(\alpha Q) |\bub;\bvb \rangle = \sum\varkappa_1^{-a_1}\varkappa_2^{b_1} r_1^{(1)}(\buc_{\st})r_1^{(2)}(\bub_{\so})r_3^{(2)}(\bvc_{\so})r^{(1)}_3(\bvb_{\st})f(\bvc_{\st},\buc_{\so})f(\bvb_{\so},\bub_{\st})\\
\times S_1\left(\buc_{\so};\bvc_{\so}|\bub_{\so};\bvb_{\so}\right)S_2\left(\buc_{\st};\bvc_{\st}|\bub_{\st};\bvb_{\st}\right)
f(\buc_{\so},\buc_{\st})f(\bub_{\st},\bub_{\so})f(\bvc_{\so},\bvc_{\st})f(\bvb_{\st},\bvb_{\so}),
\end{multline}
where $r_i^{(\ell)}$, $\ell=1,2, i=1,3$ are partial $r_i$ of $\ell$-subsystem and $S_{\ell}$, $\ell=1,2$ are scalar products of the partial Bethe vectors, $\varkappa_i=e^{\alpha_i}$, $i=1,2$ and partitions are the same as above. Operator $\exp(\alpha Q)$ only contributes producing  the additional factors $\varkappa^{-a_1}_1$, $\varkappa_2^{b_1}$, where $a_1$, $b_1$ are the numbers of particles of the first and the second type in the first subsystem, i.e. eigenvalues of $Q_1^{(1)}$ and $Q_2^{(1)}$.

Substituting \eqref{scalp} into \eqref{coprod}  we get
\be{sum_form1}
\begin{split}
\langle \buc;\bvc |\exp(\alpha Q)& |\bub;\bvb \rangle =\sum \varkappa_1^{-a_1}\varkappa_2^{b_1} r_1^{(1)}(\buc_{\st})r^{(2)}_1(\bub_{\so})r_3^{(2)}(\bvc_{\so})r_3^{(1)}(\bvb_{\st})\\
\times &f(\bvc_{\st},\buc_{\so})f(\bvb_{\so},\bub_{\st})f(\buc_{\so},\buc_{\st})f(\bub_{\st},\bub_{\so})f(\bvc_{\so},\bvc_{\st})f(\bvb_{\st},\bvb_{\so})\\
\times r_1^{(1)}&(\bub_1)r_1^{(1)}(\buc_2)r_3^{(1)}(\bvb_1)r^{(1)}_3(\bvc_2)f(\buc_1,\buc_2)f(\bub_2,\bub_1)f(\bvc_2,\bvc_1)f(\bvb_1,\bvb_2)\\
\times &f(\bvc_1,\buc_1)f(\bvb_2,\bub_2){ Z_{a_1-k_1,n_1}(\buc_2;\bub_2|\bvc_1;\bvb_1)}{ Z_{k_1,b_1-n_1}(\bub_1;\buc_1|\bvb_2,\bvc_2)}\\
\times r_1^{(2)}&(\bub_3)r_1^{(2)}(\buc_4)r_3^{(2)}(\bvb_3)r_3^{(2)}(\bvc_4)f(\buc_3,\buc_4)f(\bub_4,\bub_3)f(\bvc_4,\bvc_3)f(\bvb_3,\bvb_4)\\
\times &f(\bvc_3,\buc_3)f(\bvb_4,\bub_4){ Z_{a_2-k_2,n_2}(\buc_4;\bub_4|\bvc_3;\bvb_3)}{ Z_{k_2,b_2-n_2}(\bub_3;\buc_3|\bvb_4;\bvc_4)},
\end{split}
\ee
here the sets are divided as $\bub_{\so}\to\{\bub_1,\bub_2\}$, $\bub_{\st}\to\{\bub_3,\bub_4\}$, $\bvb_{\so}\to\{\bvb_1,\bvb_2\}$, $\bvb_{\st}\to\{\bvb_3,\bvb_4\}$ and the same for $\{\buc_{\so}, \buc_{\st}, \bvc_{\so}, \bvc_{\st}\}$.

Now we want to regroup the factors under the sum over partitions in order to find the new highest coefficients (HC) for the expectation value \eqref{coprod} explicitly and derive the Reshetikhin-like formula structure. We gather the first $Z_{a,b}$ in \eqref{sum_form1} with the last and the second $Z_{a,b}$ with the third and collect all factors that depend on the same variables as corresponding $Z_{a,b}$.

Let us make some simplification of the long factors separately.
\be{vc_new}
\begin{split}
f(\bvc_1,\bvc_3)f(\bvc_2,\bvc_3)&f(\bvc_1,\bvc_4)f(\bvc_2,\bvc_4)f(\bvc_2,\bvc_1)f(\bvc_4,\bvc_3)\\
&=f(\bvc_2,\bvc_3)f(\bvc_1,\bvc_4)f(\bvc_{14},\bvc_3)f(\bvc_2,\bvc_{14}).
\end{split}
\ee
Here we denote $f(\bar y_{i,j},\bar x)=f(\bar y_i,\bar x)f(\bar y_j,\bar x)$. In the similar way we obtain
\be{uc}
\begin{split}
f(\buc_1,\buc_3)f(\buc_2,\buc_3)&f(\buc_1,\buc_4)f(\buc_2,\buc_4)f(\buc_1,\buc_2)f(\buc_3,\buc_4)\\
&=f(\buc_2,\buc_3)f(\buc_1,\buc_4)f(\buc_1,\buc_{23})f(\buc_{23},\buc_4),
\end{split}
\ee
\be{vcuc}
\begin{split}
f(\bvc_3,\buc_1)f(\bvc_4,\buc_1)&f(\bvc_3,\buc_2)f(\bvc_4,\buc_2)f(\bvc_1,\buc_1)f(\bvc_3,\buc_3)\\
&=f(\bvc_3,\buc_{23})f(\bvc_{14},\buc_1)f(\bvc_3,\buc_1)f(\bvc_4,\buc_2).
\end{split}
\ee
Using equations \eqref{vc_new}--\eqref{vcuc} and absolutely similar simplification for sets $\{\bvb,\bub\}$ we arrive at
\be{sum_form_off_shell}
\begin{split}
\langle \buc;\bvc |\exp(\alpha Q) |\bub;\bvb \rangle=\sum  r_1^{(1)}&(\buc_3)r_1^{(1)}(\buc_4)r_1^{(2)}(\bub_1)r_1^{(2)}(\bub_2)r_3^{(2)}(\bvc_1)r_3^{(2)}(\bvc_2)\varkappa_1^{-a_1}\varkappa_2^{b_1}\\
\times r_3^{(1)}(\bvb_3)r_3^{(1)}(\bvb_4) r_1^{(1)}(\bub_1)r_1^{(1)}(\buc_2)&r_3^{(1)}(\bvb_1)r_3^{(1)}(\bvc_2)r_1^{(2)}(\bub_3)r_1^{(2)}(\buc_4)r_3^{(2)}(\bvb_3)r_3^{(2)}(\bvc_4)\\
\times&\left[Z_{a_1-k_1,n_1}(\buc_2;\bub_2|\bvc_1;\bvb_1)Z_{k_2,b_2-n_2}(\bub_3;\buc_3|\bvb_4;\bvc_4)\right]\\
\times&\left[Z_{k_1,b_1-n_1}(\bub_1;\buc_1|\bvb_2;\bvc_2)Z_{a_2-k_2,n_2}(\buc_4;\bub_4|\bvc_3;\bvb_3)\right]\\
\times f(\buc_2,\buc_3) f(\buc_1,\buc_4)f(\buc_1,\buc_{23})&f(\buc_{23},\buc_4) f(\bub_4,\bub_1)f(\bub_3,\bub_2) f(\bub_{23},\bub_1)f(\bub_4,\bub_{23})\\
\times  f(\bvc_2,\bvc_3)f(\bvc_1,\bvc_4) f(\bvc_{14},\bvc_3)&f(\bvc_2,\bvc_{14})f(\bvb_4,\bvb_1) f(\bvb_3,\bvb_2)f(\bvb_{14},\bvb_2)f(\bvb_3,\bvb_{14})\\
\times  f(\bvc_3,\buc_{23})f(\bvc_{14},\buc_1)f(\bvc_3,\buc_1)&f(\bvc_4,\buc_2) f(\bvb_2,\bub_{23})f(\bvb_{14},\bub_4)f(\bvb_1,\bub_3) f(\bvb_2,\bub_4).
\end{split}
\ee
Regrouping now the factors in \eqref{sum_form_off_shell}, we can express $\langle \buc;\bvc |\exp\left(\alpha Q\right)|\bub;\bvb \rangle$  in terms of two new highest coefficients (factors in the first and the second brackets correspondingly)
\begin{multline}
\label{sum_form_off_shell_2}
\langle \buc;\bvc |\exp(\alpha Q) |\bub;\bvb \rangle=\sum r_1^{(1)}(\buc_{23})r_3^{(1)}(\bvb_{14})r_1^{(2)}(\bub_{23})r_3^{(2)}(\bvc_{14})\varkappa_1^{k_2-a_2-k_1}\varkappa_2^{n_1-n_2+b_2}\\
\times \left\{\sum r_1(\buc_4)r_1(\bub_1)r_3(\bvc_2)r_3(\bvb_3) Z_{k_1,b_1-n_1}(\bub_1;\buc_1|\bvb_2;\bvc_2)Z_{a_2-k_2,n_2}(\buc_4;\bub_4|\bvc_3;\bvb_3)\right.\\
\times \varkappa_1^{a_2-k_2}\varkappa_2^{b_1-n_1}  
f(\buc_1,\buc_4)f(\buc_1,\buc_{23})f(\buc_{23},\buc_4)f(\bub_4,\bub_1)f(\bub_{23},\bub_1)f(\bub_4,\bub_{23})\\
\times f(\bvc_2,\bvc_3)f(\bvc_{14},\bvc_3)f(\bvc_2,\bvc_{14})f(\bvb_3,\bvb_2)f(\bvb_{14},\bvb_2)f(\bvb_3,\bvb_{14})\\
 \left. \frac{}{}\times f(\bvc_3,\buc_{23})f(\bvc_{14},\buc_1)f(\bvc_3,\buc_1)f(\bvb_2,\bub_{23})f(\bvb_{14},\bub_4)f(\bvb_2,\bub_4)\right\}\\
\times \left\{\sum \varkappa_1^{k_1-a_1}\varkappa_2^{n_2-b_2}Z_{a_1-k_1,n_1}(\buc_2;\bub_2|\bvc_1;\bvb_1)Z_{k_2,b_2-n_2}(\bub_3;\buc_3|\bvb_4;\bvc_4)\right.\\
\times \left. f(\buc_2,\buc_3)f(\bub_3,\bub_2)f(\bvc_1,\bvc_4)f(\bvb_4,\bvb_1)f(\bvc_4,\buc_2)f(\bvb_1,\bub_3)\frac{}{}\right\}.
\end{multline}
The common summation is taken over partitions $\buc\to \{\buc_{14},\buc_{23}\}$, $\bvc\to \{\bvc_{14},\bvc_{23}\}$ and the same for $\{\bub,\bvb\}$. In the first brackets summation is taken over partitions $\buc_{14}\to\{\buc_1,\buc_4\}$, $\bvc_{23}\to\{\bvc_2,\bvc_3\}$ and the same for $\{\bub,\bvb\}$. In the second brackets summation is taken over partitions $\buc_{23}\to\{\buc_2,\buc_3\}$, $\bvc_{14}\to\{\bvc_1,\bvc_4\}$ and the same for $\{\bub,\bvb\}$.

Let us make further simplifications now. After applying  Bethe equations \eqref{BEgl3} for $r_1(\bub_1)$ and $r_3(\bvb_3)$  the long factors in the first HC (the first brackets in \eqref{sum_form_off_shell_2}) can be simplified. 
\begin{multline}
\label{HC1new}
\sum f(\buc_1,\buc_4)f(\buc_1,\buc_{23})f(\buc_{23},\buc_4)\frac{f(\bub_1,\bub_{234})}{f(\bub_{234},\bub_1)}f(\bub_{234},\bub_1)f(\bub_4,\bub_{23})\\
\times r_1(\buc_4)r_3(\bvc_2)f(\bvc_2,\bvc_3)f(\bvc_{14},\bvc_3)f(\bvc_2,\bvc_{14})f(\bvb_3,\bvb_{124})f(\bvb_{14},\bvb_2)\frac{f(\bvb_{124},\bvb_3)}{f(\bvb_3,\bvb_{124})}\\
\times f(\bvc_3,\buc_{23})f(\bvc_{14},\buc_1)f(\bvc_3,\buc_1)f(\bvb_{23},\bub_{14})f(\bvb_{23},\bub_{23})f(\bvb_{14},\bub_{14})f(\bvb_3,\bub_1)\\
=f(\bub_{14},\bub_{23})f(\buc_{14},\buc_{23})f(\bvc_{14},\bvc_{23})f(\bvb_{14},\bvb_{23})f(\bvb_{23},\bub)f(\bvb_{14},\bub_{14})\\
\times \sum  
 f(\buc_1,\buc_4)\frac{f(\buc_{23},\buc_4)}{f(\buc_4,\buc_{23})}f(\bub_1,\bub_4)f(\bvc_2,\bvc_3)\frac{f(\bvc_2,\bvc_{14})}{f(\bvc_{14},\bvc_2)}f(\bvb_2,\bvb_3)\\
\times r_1(\buc_4)r_3(\bvc_2)\frac{f(\bvc_{14},\buc_{14})f(\bvc_{23},\buc_{23})}{f(\bvc_{14},\buc_4)f(\bvc_2,\buc_{23})}f(\bvc_3,\buc_1)f(\bvb_3,\bub_1).
\end{multline}
The sum is taken over partitions $\bu_{23}\to\{\bu_2,\bu_3\}$, $\bv_{14}\to\{\bv_1,\bv_4\}$. The factors with $f(\bvb_{\alpha},\bub_{\beta})$ here were simplified in the following way:
\be{factor_vb_ub}
\begin{split}
f(\bvb_2,\bub_{23})&f(\bvb_{14},\bub_4)f(\bvb_2,\bub_4)f(\bvb_3,\bub)f(\bvb,\bub_1)\\
&=f(\bvb_{23},\bub_{14})f(\bvb_{23},\bub_{23})f(\bvb_{14},\bub_{14})f(\bvb_3,\bub_1).
\end{split}
\ee
Now we absorb part of factors in the coefficients $r_1$, $r_3$
\be{r_scaling}
\begin{split}
r_1(\buc_4)\longrightarrow \hat r_1(\buc_4)=\frac{r_1(\buc_4)}{f(\bvc_{14},\buc_4)}\frac{f(\buc_{23},\buc_4)}{f(\buc_4,\buc_{23})},\\
r_3(\bvc_2)\longrightarrow\hat r_3(\bvc_2)=\frac{r_3(\bvc_2)}{f(\bvc_2,\buc_{23})}\frac{f(\bvc_2,\bvc_{14})}{f(\bvc_{14},\bvc_2)}.
\end{split}
\ee
Finally, \eqref{HC1new} can be written as
\be{HC1new_final}
\begin{split}
&f(\bub_{14},\bub_{23})f(\buc_{14},\buc_{23})f(\bvc_{14},\bvc_{23})f(\bvb_{14},\bvb_{23})f(\bvb_{23},\bub_{23})f(\bvb_{14},\bub_{14})f(\bvc_{23},\buc_{23})f(\bvc_{14},\buc_{14})\\
&\times f(\bvb_{23},\bub_{14})\sum \varkappa_1^{a_2-k_2}\varkappa_2^{b_1-n_1}\hat r_1(\buc_4)\hat r_3(\bvc_2) f(\buc_1,\buc_4)f(\bub_1,\bub_4)f(\bvc_2,\bvc_3)f(\bvb_2,\bvb_3)\\
&\times f(\bvb_3,\bub_1)f(\bvc_3,\buc_1) Z_{k_1,b_1-n_1}\left(\bub_1;\buc_1|\bvb_2;\bvc_2\right)Z_{a_2-k_2,n_2}\left(\buc_4;\bub_4|\bvc_3;\bvb_3\right).
\end{split}
\ee
The last two lines  of \eqref{HC1new_final} coincide with the scalar product \eqref{scalp} where rapidities $\{\bub_{14},\bvb_{23}\}$ satisfy (untwisted) Bethe equations, $\{\buc_{14},\bvc_{23}\}$ are arbitrary and $\{r_1(u),r_3(v)\}$ are modified according to \eqref{r_scaling} and posses twists: $r_1\to \varkappa_1\hat r_1$, $r_3\to\varkappa_2\hat r_3$. Partitions are $\bu_{23}\to\{\bu_2,\bu_3\}$, $\bv_{14}\to\{\bv_1,\bv_4\}$. We denote this scalar product with the additional normalisation $(f(\bvc_{23},\buc_{14})f(\bvb_{23},\bub_{14}))^{-1}$ as $\hat\Theta_{a_2-k_2+k_1,b_1-n_1+n_2}^{\alpha}\left(\buc_{14},\bub_{14}|\bvb_{23},\bvc_{23}\right)$.

The second highest coefficient (second brackets in \eqref{sum_form_off_shell_2}) coincides with a scalar product of the on-shell Bethe vector with parameters $\{\bub_{23},\bvb_{14}\}$ and the twisted-on-shell Bethe vector with parameters $\{\buc_{23},\bvc_{14}\}$ normalised  by factor $\left(f(\bvb_{14},\bub_{23})f(\bvc_{14},\buc_{23})\right)^{-1}$. We denote this normalised scalar product  as $\Theta_{a_1-k_1+k_2,b_2-n_2+n_1}^{\alpha}\left(\buc_{23},\bub_{23}|\bvb_{14},\bvc_{14}\right)$.

Let us rename for brevity the sets from \eqref{sum_form_off_shell_2} as $\bu_{23}\to\bu_2$, $\bu_{14}\to\bu_1$, $\bv_{23}\to\bv_2$ and $\bv_{14}\to\bv_1$,  and cardinalities of sets are also relabeled as $a_1-k_1+k_2\to a_2$, $a_2+k_1-k_2\to a_1$, $b_1-n_1+n_2\to b_2$ and $b_2-n_2+n_1\to b_1$. Then the matrix element of the operator $\exp\left(\alpha Q_1\right)$ between off-shell and on-shell Bethe vectors can be presented in form of Reshetikhin sum formula where each set is divided into two subsets and the new highest coefficients $\Theta$ and $\hat\Theta$ are given by the scalar products of Bethe vectors
\be{operator_Reshetikhin_gl3}
\begin{split}
\langle \buc;\bvc |\exp(\alpha Q)& |\bub;\bvb \rangle=\sum \varkappa_1^{a_2}\varkappa_2^{b_1} r_1^{(1)}(\buc_2)r_3^{(1)}(\bvb_1)r_1^{(2)}(\bub_2)r_3^{(2)}(\bvc_1)\\
\times f(\bub_1,\bub_2)&f(\buc_1,\buc_2)f(\bvc_1,\bvc_2)f(\bvb_1,\bvb_2)f(\bvb_2,\bub_2)f(\bvb_1,\bub_1)f(\bvc_2,\buc_2)f(\bvc_1,\buc_1)\\
&\times f(\bvb_2,\bub_1)f(\bvc_2,\buc_1)\hat\Theta_{a_1,b_2}^{\alpha}\left(\buc_1,\bub_1|\bvb_2,\bvc_2\right)\Theta_{a_2,b_1}^{\alpha}\left(\buc_2,\bub_2|\bvb_1,\bvc_1\right). 
\end{split}
\ee
Here $\bub\to\{\bub_1,\bub_2\}$, $\bvb\to\{\bvb_1,\bvb_2\}$ and the same for sets $\{\buc,\bvc\}$. This formula is a direct $\mathfrak{gl}{(3)}$ analogue of one derived in \cite{ME} for the algebra symmetry $\mathfrak{gl}(2)$ related models (see formula C.7 there)\footnote{Particular case of this formula was also derived in algebra symmetry $\mathfrak{gl}(2)$ related models in \cite{Izergin1984}, but the set $\buc$, was also on-shell there.}.  It is easy to check that the similar formula holds for the algebra symmetry $\mathfrak{gl}(2|1)$ related models. The  only difference is replacement of factors $f(\bvb_1,\bvb_2)f(\bvc_1,\bvc_2)$ by $g(\bvb_1,\bvb_2)g(\bvc_1,\bvc_2)$ and disappearance of the factor $f(\bvc_2,\bvc_{14})/f(\bvc_{14},\bvc_2)$ in the second line of \eqref{r_scaling}.
\be{operator_Reshetikhin_gl21}
\begin{split}
\langle \buc;\bvc |\exp(\alpha Q)& |\bub;\bvb \rangle=\sum \varkappa_1^{a_2}\varkappa_2^{b_1} r_1^{(1)}(\buc_2)r_3^{(1)}(\bvb_1)r_1^{(2)}(\bub_2)r_3^{(2)}(\bvc_1)\\
\times f(\bub_1,\bub_2)&f(\buc_1,\buc_2)g(\bvc_1,\bvc_2)g(\bvb_1,\bvb_2)f(\bvb_2,\bub_2)f(\bvb_1,\bub_1)f(\bvc_2,\buc_2)f(\bvc_1,\buc_1)\\
&\times f(\bvb_2,\bub_1)f(\bvc_2,\buc_1)\hat\Theta_{a_1,b_2}^{\alpha}\left(\buc_1,\bub_1|\bvb_2,\bvc_2\right)\Theta_{a_2,b_1}^{\alpha}\left(\buc_2,\bub_2|\bvb_1,\bvc_1\right). 
\end{split}
\ee

\section{Integral representation for form factors \label{me_integral}}

Representations \eqref{operator_Reshetikhin_gl3}-\eqref{operator_Reshetikhin_gl21} allow further simplification in case if explicit compact formulae for $\Theta_{a,b}$ and $\hat\Theta_{a,b}$ are known. The compact formulae (determinant representations) for the the $\Theta^{\alpha}_{m,n}$, $\hat\Theta^{\alpha}_{m,n}$ are known for algebra symmetry $\mathfrak{gl}(2|1)$ related models. For $\Theta^{\alpha}_{m,n}$ it was derived in \cite{SLHRP3} while the scalar product of the off-shell and on-shell Bethe vectors that coincides with $\hat\Theta^{\alpha}_{m,n}$ was derived in \cite{Liashyk}. Further in this paper we restrict ourselves by the algebra symmetry $\mathfrak{gl}(2|1)$ case. Our goal now is a derivation from \eqref{operator_Reshetikhin_gl21} integral representation suitable for the asymptotic analysis.

Explicit determinant representations of $\Theta^{\alpha}_{a,b}$ is given by
\be{tONS-ONS}
\begin{split}
\Theta_{a,b}^{\alpha}\left(\bar z,\bub|\bvb,\bar y\right)=g(\bub,\by)\Delta_g(\bub)\Delta_g(\by)\Delta'_g(\bar z)\Delta'_g(\bar y)h(\bub,\bub)h(\by,\bub)\det_{a,b} \mathcal N,
\end{split}
\ee
where diagonal parts of block matrix $\mathcal N\left(\bub,\bvb|\bz,\by\right)$ are defined as
\be{N}
\begin{split}
&\mathcal N_{11}=t(z_j,\ub_k)\frac{f(\bvb,\ub_k)h(\bar z,\ub_k)}{f(\bar y,\ub_k)h(\bub,\ub_k)}+\frac{1}{\varkappa_1} t(\ub_k,z_j)\frac{h(\ub_k,\bar z)}{h(\ub_k,\bub)},\quad j=1,\dots,a,\quad k=1,\dots,a,\\
&\mathcal N_{22}=\delta_{jk}\frac{g(y_k,\bvb)}{g(y_k,\bar y_k)}\left(1-\frac{1}{\varkappa_2}\frac{f(y_k,\bar z)}{f(y_k,\bub)}\right),\qquad j=1,\dots,b,\qquad k=1,\dots,b,
\end{split}
\ee
and antidiagonal as
\be{N_anti_diag}
\begin{split}
&\mathcal N_{12}=\frac{1}{\varkappa_{1}} t(y_k,z_j)\frac{h(\ub_k,\bar z)}{h(\ub_k,\bub)},\qquad j=1,\dots,a,\qquad k=1,\dots,b,\\
&\mathcal N_{21}=\frac{g(\ub_k,\bvb)}{g(\ub_k,\bar y)}\left(g(\ub_k,y_j)+\frac{\varkappa_1/\varkappa_2}{h(y_j,\ub_k)}\right), \qquad j=1,\dots,b,\qquad k=1,\dots,a.\\
\end{split}
\ee
We can introduce two functions $\mathcal Y_{\alpha}^{(i)}$, $i=1,2$ that are defines the twisted Bethe equations
\be{BE12}
\begin{split}
&\mathcal Y^{(1)}_{\alpha}\left(z_j|\bar z,\bar y\right)=1-\varkappa_1\frac{r_1(z_j)}{f(\bar y,z_j)}
\frac{f(\bar z_j,z_j)}{f(z_j,\bar z_j)},\\
&\mathcal Y^{(2)}_{\alpha}\left(y_j|\bar z,\bar y\right)=1-\varkappa_2\frac{r_3(y_j)}{f(y_j,\bar z)}.
\end{split}
\ee
It is easy to check that
\be{res_N}
\begin{split}
&\Res_{z_j=\ub_j}\frac{\det\limits_{a,b}\mathcal N\left(\bub,\bvb|\bz,\by\right)}{\mathcal Y_{\alpha}^{(1)}\left(z_j|\bz,\by\right)}=-\frac{1}{\varkappa_1}\frac{h(\ub_j,\bz_j)}{h(\ub_j,\bub_j)}\det_{a-1,b}\mathcal N\left(\bub_j,\bvb|\bz_j,\by\right),\\
&\Res_{y_j=\vb_j}\frac{\det\limits_{a,b}\mathcal N\left(\bub,\bvb|\bz,\by\right)}{\mathcal Y_{\alpha}^{(2)}\left(y_j|\bz,\by\right)}=-\frac{1}{\varkappa_2}\frac{g(\vb_j,\bvb_j)}{g(\vb_j,\by_j)}\frac{f(\vb_j,\bz)}{f(\vb_j,\bub)}\det_{a,b-1}\mathcal N\left(\bub,\bvb_j|\bz,\by_j\right).
\end{split}
\ee

The second scalar product is given by
\be{tONS-OFS}
\begin{split}
\hat\Theta_{a,b}^{\alpha}\left(\buc,\bz|\by,\bvc\right)=g(\bvc,\bz)h(\bvc,\bz)h(\bz,\bz)\Delta'_g(\buc)\Delta_g(\bz)\Delta'_g(\bvc)\Delta_g(\bvc)\det_{a,b} \mathcal M,
\end{split}
\ee
where diagonal  blocks of matrix $\mathcal M\left(\bz,\by|\buc,\bvc\right)$ are defined as
\be{M}
\begin{split}
&\mathcal M_{11}=\left(\phi(\uc_j)g(z_k,\uc_j)-\frac{1}{h(z_k,\uc_j)}\right)\frac{h(z_k,\buc)}{h(z_k,\bz)}+\left(g(\uc_j,z_k)-\frac{\phi(\uc_j)}{h(\uc_j,z_k)}\right)\frac{f(\by,z_k)h(\buc,z_k)}{f(\bvc,z_k)h(\bz,z_k)},\\ 
&\qquad j=1,\dots,a,\qquad k=1,\dots,a,\\
&\mathcal M_{22}=\delta_{jk}\left(1-\varkappa_2\frac{r_3(\vc_j)}{f(\vc_j,\bz)}\right)\frac{g(\vc_j,\by)}{g(\vc_j,\bvc_j)},\qquad j=1,\dots,b,\qquad k=1,\dots,b,
\end{split}
\ee
and antidiagonal blocks are
\be{M_anti_diag}
\begin{split}
&\mathcal M_{12}=\left(-\phi(\uc_j)g(\vc_k,\uc_j)+\frac{1}{h(\vc_k,\uc_j)}\right)\frac{h(\vc_k,\buc)}{h(\vc_k,\bz)},\qquad j=1,\dots,a,\qquad k=1,\dots,b,\\
&\mathcal M_{21}=\left(g(\vc_j,z_k)-\varkappa_2\frac{r_3(\vc_j)}{f(\vc_j,\buc)h(\vc_j,z_k)}\right)\frac{g(\by,z_k)}{g(\bvc,z_k)},\qquad j=1,\dots,b,\qquad k=1,\dots,a,
\end{split}
\ee
with
\be{phi}
\phi(u_j)=\varkappa_1\frac{r_1(u_j)}{f(\bv,u_j)}\frac{f(\bu_j,u_j)}{f(u_j,\bu_j)}.
\ee
It is easy to check that
\be{res_M}
\begin{split}
&\Res_{z_j=\uc_j}\frac{\det\limits_{a,b}\mathcal M\left(\bz,\by|\buc,\bvc\right)}{\mathcal Y_{\alpha}^{(1)}\left(z_j|\bz,\by\right)}=-\frac{f(\by,\uc_j)}{f(\bvc,\uc_j)}\frac{h(\buc,\uc_j)}{h(\bz,\uc_j)}\det_{a-1,b}\mathcal M\left(\bz_j,\by|\buc_j,\bvc\right)_{mod},\\
&\Res_{y_j=\vc_j}\frac{\det\limits_{a,b}\mathcal M\left(\bz,\by|\buc,\bvc\right)}{\mathcal Y_{\alpha}^{(2)}\left(y_j|\bz,\by\right)}=-\frac{g(\vc_j,\by)}{g(\vc_j,\bvc)}\det_{a,b-1}\mathcal M\left(\bz,\by_j|\buc,\bvc_j\right)_{mod},
\end{split}
\ee
where in the r.h.s. the modification \eqref{r_scaling} of $r_1(\buc)$, $r_3(\bvc)$  is taken.

Substituting \eqref{tONS-ONS}, \eqref{tONS-OFS} to \eqref{operator_Reshetikhin_gl21} we arrive at
\be{operator_Reshetikhin_2}
\begin{split}
&\langle \buc;\bvc |\exp(\alpha Q_1) |\bub;\bvb \rangle=f(\bvb,\bub)\sum \varkappa_1^{a_2}\varkappa_2^{b_1}\frac{r_1^{(1)}(\buc_2)r_3^{(2)}(\bvc_1)}{r_1^{(1)}(\bub_2)r_3^{(2)}(\bvb_1)}\\
&\times f(\bub_2,\bub_1)f(\buc_1,\buc_2)g(\bvc_1,\bvc_2)g(\bvb_1,\bvb_2)f(\bvb_2,\bub_2)f(\bvb_1,\bub_1)f(\bvc_2,\buc_2)f(\bvc_1,\buc_1)\\
&\times f(\bvb_1,\bub_2)f(\bvc_2,\buc_1)\left[f(\bvc_1,\bub_2)\Delta_g(\bvc_1)\Delta'_g(\bvc_1)\Delta_g(\bub_2)\Delta'_g(\buc_2)\right][h(\bub_1,\bub_1)h(\bub_2,\bub_2)]\\
&\times \left[f(\bvc_2,\bub_1)\Delta'_g(\buc_1)\Delta_g(\bub_1)\Delta_g(\bvc_2)\Delta'_g(\bvc_2)\right]\det_{a_2,b_1}\mathcal N\left(\bub_2,\bvb_1|\buc_2,\bvc_1\right)\det_{a_1,b_2}\mathcal M\left(\bub_1,\bvb_2|\buc_1,\bvc_2\right). 
\end{split}
\ee
Here $\bub\to\{\bub_1,\bub_2\}$, $\bvb\to\{\bvb_1,\bvb_2\}$ and the same for sets $\{\buc,\bvc\}$.

Now we are in position to formulate one of the main results of the paper.
\begin{lemma}
The sum over partitions \eqref{operator_Reshetikhin_2} can be expressed via the multiple contour integral
\be{contour_corr}
\begin{split}
\langle \buc;\bvc |\exp(\alpha Q)& |\bub;\bvb \rangle
=\frac{1}{a!b!}\oint\limits_{\buc\cup\bub}\prod_{j=1}^a\frac{dz_j}{2\pi i}\oint\limits_{\bvc\cup\bvb}\prod_{j=1}^b\frac{d y_j}{2\pi i}\\
&\times\frac{r_1^{(1)}(\bz)r_3^{(2)}(\by)}{r_1^{(1)}(\bub)r_3^{(2)}(\bvb)}\frac{\det\limits_{a,b}\mathcal N\left(\bar z,\bub|\bvb,\bar y\right)\det\limits_{a,b}\mathcal M\left(\buc,\bar z|\bar y,\bvc\right)}{\prod_{j=1}^a \mathcal Y^{(1)}_{\alpha}\left(z_j|\bar z,\bar y\right)\prod_{j=1}^b\mathcal Y^{(2)}_{\alpha}\left(y_j|\bar z,\bar y\right)}S.
\end{split}
\ee
Here contours encircle points $\buc$ and $\bub$ for $\bz$ and do not include any other singularities of the integrand  and around $\bvc$, $\bvb$ for $\by$. Factor $S=S(\bz,\by|\bub,\buc;\bvb,\bvc)$ is defined as $S=S_1S_2S_3$ with
\be{S3}
\begin{split}
&S_1=h(\bub,\bub),\qquad\qquad
S_2=\Delta_g^2(\bvc)\Delta_g(\buc)\Delta_g(\bub),\\
&S_3(\bz;\by)=f(\bvc,\buc)f(\bvb,\bub)\frac{f(\by,\bub)f(\bvc,\bz)}{f(\by,\bz)f(\by,\bz)}.
\end{split}
\ee
\end{lemma}
{\it Proof.} In order to prove the lemma it is enough to calculate integrals explicitly. We should take into account that we can not compute residues at $z_j=u_k$ and $z_i=u_k$ for $j\ne k$ (and the same for $y$ variables).  Taking residues according to \eqref{res_N}, \eqref{res_M} we arrive at four sums over partitions $\buc\rightarrow\{\buc_1,\buc_2\}$, $\bub\rightarrow\{\bub_1,\bub_2\}$, $\bvc\rightarrow\{\bvc_1,\bvc_2\}$, $\bvb\rightarrow\{\bvb_1,\bvb_2\}$
\be{integral_1}
\begin{split}
\frac{1}{b!}\sum_{\buc\rightarrow\{\buc_1,\buc_2\}}&\sum_{\bub\rightarrow\{\bub_1,\bub_2\}}\oint\frac{d\by}{(2\pi i)^b}\frac{\det_{a_2,b} \mathcal N\left(\bub;\bvb|\bz;\by\right)\det_{a_1,b}\mathcal M\left(\bz;\by|\buc;\bvc\right)}{\prod\limits_{j=1}^b\mathcal Y^{(2)}_{\varkappa}\left(y_j|\bz,\by\right)}\\
&\times \left. S\frac{1}{\varkappa_{1}^{a_1}}\frac{r_1^{(1)}(\bz)r_3^{(2)}(\by)}{r_1^{(1)}(\bub)r_3^{(2)}(\bvb)}\frac{h(\bub_1,\bz)}{h(\bub_1,\bub)}\frac{f(\by,\buc_2)h(\buc,\buc_2)}{f(\bvc,\buc_2)h(\bz,\buc_2)}\right|_{\bz=\{\bub_1,\buc_2\}}
\end{split}
\ee
\be{integral_2}
\begin{split}
&=\sum_{\substack{\buc\rightarrow\{\buc_1,\buc_2\}\\\bub\rightarrow\{\bub_1,\bub_2\} }}\sum_{\substack{\bvc\rightarrow\{\bvc_1,\bvc_2\}\\ \bvb\rightarrow\{\bvb_1,\bvb_2\}}}\det_{a_2,b_1}\mathcal N\left(\bub;\bvb|\bz;\by\right)\det_{a_1,b_2}\mathcal M\left(\bz;\by|\buc;\bvc\right)\frac{S}{\varkappa_{1}^{a_1}\varkappa_{2}^{b_2}}\\ 
&\times\left. \frac{r_1^{(1)}(\bz)r_3^{(2)}(\by)}{r_1^{(1)}(\bub)r_3^{(2)}(\bvb)}\frac{h(\bub_1,\bz)}{h(\bub_1,\bub)}\frac{f(\by,\buc_2)h(\buc,\buc_2)}{f(\bvc,\buc_2)h(\bz,\buc_2)}\frac{g(\bvb_2,\bvb)f(\bvb_2,\bz)}{g(\bvb_2,\by)f(\bvb_2,\bub)}\frac{g(\bvb_1,\by)}{g(\bvc_1,\bvc)}\right|_{\substack{\bz=\{\bub_1,\buc_2\},\\ \by=\{\bvc_1,\bvb_2\}} }.
\end{split}
\ee
After the substitution $\bz=\{\bub_1,\buc_2\}$, $\by=\{\bvc_1,\bvb_2\}$ from  the explicit form of determinants \eqref{N}, \eqref{M} it is clear, that they will be reduced to  the scalar products of the on-shell and twisted-on-shell Bethe vectors with sets $\{\bu_2,\bvb_1\}$, $\{\buc_2,\bvc_1\}$ and the scalar product of the on-shell and off-shell Bethe vectors with sets $\{\buc_1,\bvc_2\}$, $\{\bub_1,\bvb_2\}$ correspondingly and modified $\hat r_1, \hat r_3$, as it should be in \eqref{operator_Reshetikhin_2}. It remains to check all other factors in \eqref{operator_Reshetikhin_2}.
\be{h_factor}
\begin{split}
&\left.\frac{h(\bub_1,\bz)}{h(\bub_1,\bub)}\frac{h(\buc,\buc_2)}{h(\bz,\buc_2)}h(\bub,\bub)\right|_{\bz\rightarrow\{\bub_1,\buc_2\}}\\
&=\frac{h(\bub_1,\buc_2)}{h(\bub_1,\bub_2)}\frac{h(\buc_1,\buc_2)}{h(\bub_1,\buc_2)}h(\bub_1,\bub_2)h(\bub_2,\bub_1)h(\bub_1,\bub_1)h(\bub_2,\bub_2)\\
&=h(\bub_1,\bub_1)h(\bub_2,\bub_2)h(\bub_2,\bub_1)h(\buc_1,\buc_2).
\end{split}
\ee
Simplify part of factors in \eqref{operator_Reshetikhin_2} that contains $g(\bx,\by)$ and $\Delta_g(\bx)$ (here $\bx, \by=\bub, \buc$  or $\bvc, \bvb$)
\be{g_factor_lhs}
\begin{split}
\Delta_g(\bvc_1)\Delta_g'(\bvc_1)&\Delta_g(\bub_2)\Delta_g(\buc_2)g(\bvc_1,\bvc_2)g(\bvb_1,\bvb_2)\Delta_g(\buc_1)\Delta_g(\bub_1)\\
\times &\Delta_g(\bvc_2)\Delta_g'(\bvc_2)g(\buc_1,\buc_2)g(\bub_1,\bub_2)=\Delta^2_g(\bvc)\Delta_g(\buc)\Delta_g(\bub)\frac{g(\bvb_1,\bvb_2)}{g(\bvc_1,\bvc_2)}.
\end{split}
\ee
The last expression coincides with factors $g$ coming form residue computation and $S_2$ in \eqref{integral_2}
\be{g_factor_rhs}
S_2\frac{g(\bvb_2,\bvb_1)}{g(\bvb_2,\bvc_1)}\frac{g(\bvc_1,\bvb_2)}{g(\bvc_1,\bvc_2)}.
\ee
Finally, simplify part of factors in \eqref{operator_Reshetikhin_2} that contains $f(\bvc,\bub)$ or $f(\bvb,\buc)$
\be{f_factor_lhs}
\begin{split}
f(\bvc_1,\bub_2)f(\bvc_2,\bub_2)f(\bvb_1,\bub_2)f(\bvc_2,\buc_1)&f(\bvc_2,\buc_2)f(\bvc_1,\buc_1)f(\bvb_2,\bub_2)f(\bvb_1,\bub_1)\\
&=f(\bvc,\buc)f(\bvb,\bub)\frac{f(\bvc_1,\bub_2)f(\bvc_2,\bub_1)}{f(\bvc_1,\buc_2)f(\bvb_2,\bub_1)}.
\end{split}
\ee
The last expression coincides with factors $f$ coming from computation of residues and $S_3$ in \eqref{integral_2}
\be{f_factor_rhs}
\begin{split}
f(\bvc,\buc)f(\bvb,\bub)\frac{f(\bvb_2,\bz)f(\by,\buc_2)}{f(\bvb_2,\bub)f(\bvc,\buc_2)}& \left. \frac{f(\by,\bub)f(\bvc,\bz)}{f(\by,\bz)f(\by,\bz)}\right|_{\by\rightarrow\{\bvc_1,\bvb_2\},\; \bz\rightarrow\{\bub_1,\buc_2\}}\\
&=f(\bvc,\buc)f(\bvb,\bub)\frac{f(\bvc_1,\bub_2)f(\bvc_2,\bub_1)}{f(\bvc_1,\buc_2)f(\bvb_2,\bub_1)}.
\end{split}
\ee
Thus we convinced ourselves that the direct computation of integrals in \eqref{contour_corr} leads to \eqref{operator_Reshetikhin_2}.
\qed
In \eqref{contour_corr} $\{\buc,\bvc\}$ are still free parameters. They  may be the solution of Bethe equations including the solution $\{\bub,\bvb\}$. This allows us compute form factors and zero-temperature correlation function. This formula is a direct analogue of one derived in \cite{ME} for the algebra symmetry $\mathfrak{gl}(2)$ related models\footnote{This result was derived there even for the case of the trigonometric R-matrix.}. 

\section{Form factor series\label{F_series}}

Similarly to the algebra symmetry $\mathfrak{gl}(2)$ situation in case of fundamental model where the solution of quantum inverse problem is known $\exp(\alpha Q)$ could be expressed explicitly in terms of the (twisted) transfer matrices. Then the series \eqref{contour_corr} could be derived in a different way. The method is absolutely similar to the algebra symmetry $\mathfrak{gl}(2)$ related models, thus we give only  a very brief description here. More details are given in \cite{Q_series}.

Solution of the quantum inverse problem is formulated in the following way \cite{GohmannKorepin, Kitanine_inverted}.  Let us denote the elementary units  $(e_{ij})_{rs}=\delta_{ir}\delta_{js}$, $i,j,r,s=1,\dots,3$ acting on the site $k$ as $(e_{ij})^{(k)}$, then
\be{inverse}
(e_{ij})^{(k)}=t(c/2)^{k-1}T_{ij}(c/2)t^{-k}(c/2),
\ee
where $t(w)$ is a transfer matrix (see \ref{Bethe_ansatz}).

For the fundamental model generation function for the first subsystem consisting of only one site $m=1$ is given by
\be{Q_fundamental}
\exp(\alpha Q)=\exp\left(\alpha_1(e_{11})^{(1)}+\alpha_2(e_{33})^{(2)}\right)=\varkappa_1(e_{11})^{(1)}+(e_{22})^{(1)}-\varkappa_2(e_{33})^{(1)}.
\ee
In the case of subsystem consisting of sites $1,\dots,m$ generation function combining \eqref{inverse} and \eqref{Q_fundamental} we arrive at
\be{Q_lattice}
\exp(\alpha Q)=t_{\varkappa}^m(c/2)t^{-m}(c/2),
\ee
where $t_{\varkappa}$ is a twisted transfer matrix defined in \eqref{t_twisted} and its eigenvalue on eigenvector $|\bu,\bv\rangle$ is explicitly given by $\tau(w|\bu,\bv)$
\be{tau_explicit}
\tau(w|\bu,\bv)=\varkappa_1\lambda_1(w)f(\bu,w)+\lambda_2(w)f(w,\bu)f(\bv,w)-\varkappa_2\lambda_3(w)f(\bv,w).
\ee
For the fundamental model $\lambda_1(w)=(w+c/2)^L$ and $\lambda_2(w)=\lambda_3(w)=(w-c/2)^L$\footnote{This is the supersymmetric t-J model describing 1D hopping electron gas, see \cite{tJ, GohmannSeel} for good description.}. Momentum is given by
\be{momentum}
p(w)=\log\left(\frac{w+c/2}{w-c/2}\right).
\ee

Assuming now the completeness of the basis of twisted Bethe eigenstates\footnote{The proof of completeness requires a separate discussion, we are not aware whether such a proof was ever given rigorously for algebra symmetry $\mathfrak{gl}(2|1)$ related models.} we can insert between the transfer matrices $1=\sum_{\bar\mu,\:\bar\lambda}|\bar\mu;\bar\lambda\rangle\langle\bar\mu;\bar\lambda| \langle\bar\mu;\bar\lambda|\bar\mu;\bar\lambda\rangle^{-1}$ and obtain the following expansion for matrix elements of the generation function
\be{ff_series}
\langle\buc;\bvc|\exp\left(\alpha Q\right)|\bub;\bvb\rangle=\sum_{\bar\mu,\:\bar\lambda}\frac{\tau^m_{\varkappa}(c/2|\bar\mu;\bar\lambda)}{\tau^m(c/2|\bub;\bvb)}\frac{\langle \buc;\bvc|\bar\mu;\bar\lambda\rangle\langle\bar\mu;\bar\lambda|\bub;\bvb\rangle}{\langle\bar\mu;\bar\lambda|\bar\mu;\bar\lambda\rangle},
\ee
where $\langle\bu;\bv|\bar\mu;\bar\lambda\rangle$ denotes scalar products and we use \eqref{t_twisted}. Summation is taken over all admissible (physical) solutions\footnote{There exist also unphysical solutions of Bethe equations, where both terms in \eqref{Y_tilde} are equal to zero. It was shown in the case of algebra symmetry $\mathfrak{gl}(2)$ related model \cite{Q_series} that contributions of such solutions to the sum \eqref{ff_series} are zero. We omit here this proof since it is identical to the previous one.} of twisted Bethe equations: $\mathcal Y_{\varkappa}^{(1)}\left(\mu_i|\bar\mu;\bar\lambda\right)=0$, $i=1,\dots,a$ and $\mathcal Y_{\varkappa}^{(2)}\left(\lambda_j|\bar\mu;\bar\lambda\right)=0$, $j=1,\dots,b$.  The norm of (on-shell) Bethe vector $\langle\bar\mu;\bar\lambda|\bar\mu;\bar\lambda\rangle$ was computed in \cite{SLHRP3}. We present it here in the following form
\be{norm}
\begin{split}
\langle\bar\mu;\bar\lambda|\bar\mu;\bar\lambda\rangle=&\tilde S\det\left(\frac{\partial \tilde{\mathcal Y}_{\varkappa}^{(r)}\left(w_k^r|\bar \mu,\bar\lambda\right)}{\partial w_j^r}\right),\qquad\\
&\tilde S=\frac{\Delta_g(\bar\lambda)\Delta_g'(\bar\lambda)f(\bar\lambda,\bar\mu)
\Delta_g(\bar\mu)\Delta_g'(\bar\mu)h(\bar\mu,\bar\mu)}{\varkappa_2\lambda_3(\bar \lambda)\lambda_2(\bar\mu)h(\bar\mu,\bar\mu)f(\bar\lambda,\bar\mu)}.
\end{split}
\ee
Here, $r, s=1,2$, $w^1_j=z_j$, $j=1,\dots,a$, $w^2_j=y_j$, $j=1,\dots,b$ and
\be{Y_tilde}
\begin{split}
\tilde{\mathcal Y}_{\varkappa}^{(1)}(\mu_j)&=\varkappa_1\lambda_1(\mu_j)h(\bar\mu_j,\mu_j)+\lambda_2(\mu_j)h(\mu_j,\bar\mu_j)f(\bar\lambda,\mu_j),\\
\tilde {\mathcal Y}_{\varkappa}^{(2)}(\lambda_j)&=\varkappa_2\lambda_3(\lambda_j)+\lambda_2(\lambda_j)f(\lambda_j,\bar\mu).
\end{split}
\ee

For an arbitrary function $F(\bar\mu,\bar\lambda)$ sum can be rewritten as a contour integral around the solutions of Bethe equations using the trick
\be{sum_int}
\sum_{\bar\mu,\:\bar\lambda}F(\bar\mu,\bar \lambda)=\frac{1}{a!b!}\oint_{\bar\mu}d\bar z\oint_{\bar\lambda}d\bar y\det\left(\frac{\partial \tilde{\mathcal Y}_{\varkappa}^{(r)}(w^r_j|\bz,\by)}{\partial w^s_k}\right)\frac{ F(\bar z,\bar y)}{\prod\limits_{j=1}^a\tilde{\mathcal Y}_{\varkappa}^{(1)}(z_j|\bz;\by)\prod\limits_{j=1}^b\tilde{\mathcal Y}_{\varkappa}^{(2)}(y_j|\bz,\by)}.
\ee
Integrals on $\{\bar z,\bar y\}$ are taken around all admissible solutions $\{\bar\mu,\bar\lambda\}$ of (twisted) Bethe equations. Factorials $a!b!$ appear in order to avoid multiple counting of the Bethe states that differ only by permutations of the spectral parameters inside the set. Thus we can present \eqref{ff_series} as
\be{ff_integral}
\begin{split}
\langle \buc;\bvc |\exp(\alpha Q)& |\bub;\bvb \rangle\\=\frac{1}{a!b!}\oint\limits_{\bar\mu}&\frac{d\bar z}{(2\pi)^a}\oint_{\bar\lambda}\frac{d\bar y}{(2\pi i)^b}\frac{\tau^m_{\varkappa}(c/2|\bz;\by)}{\tau^m(c/2|\bub;\bvb)}\frac{\langle \buc;\bvc|\bz;\by\rangle\langle\bz;\by|\bub;\bvb\rangle}{\tilde S\prod\limits_{j=1}^a\tilde{\mathcal Y}_{\varkappa}^{(1)}(z_j|\bz;\by)\prod\limits_{j=1}^b\tilde{\mathcal Y}_{\varkappa}^{(2)}(y_j|\bz,\by)}.
\end{split}
\ee
Now we can substitute in \eqref{ff_integral} \eqref{tONS-ONS} for $\langle\bz;\by|\bub;\bvb\rangle$, \eqref{tONS-OFS} for $\langle\buc;\bvc|\bz;\by\rangle$ and $\tilde S$ defined in \eqref{norm}. Also we extract from $\tilde{\mathcal Y}_{\varkappa}^{(1)}(z_j|\bz,\by)$ factor $\lambda_2(z_j)h(z_j,\bz_j)f(\by,z_j)$ and from $\tilde{\mathcal Y}_{\varkappa}^{(2)}(y_j|\bz,\by)$ factor $\lambda_2(y_j)f(y_j,\bz)$, hereby $\tilde{\mathcal Y}_{\varkappa}^{(1)}\longrightarrow \mathcal Y_{\varkappa}^{(1)}$ and $\tilde{\mathcal Y}_{\varkappa}^{(2)}\longrightarrow \mathcal Y_{\varkappa}^{(2)}$. After collecting factors and elementary simplification of prefactors we arrive at
\be{contour_corr_2}
\begin{split}
\langle \buc;\bvc |\exp(\alpha Q)& |\bub;\bvb \rangle
=\frac{1}{a!b!}\oint\limits_{\bar\mu}\prod_{j=1}^a\frac{dz_j}{2\pi i}\oint\limits_{\bar\lambda}\prod_{j=1}^b\frac{dy_j}{2\pi i}\\
&\times e^{im\left(p\left(\bz\right)-p\left(\bub\right)\right)}\frac{\det\limits_{a,b}\mathcal N\left(\bar z,\bub|\bvb,\bar y\right)\det\limits_{a,b}\mathcal M\left(\buc,\bar z|\bar y,\bvc\right)}{\prod_{j=1}^a \mathcal Y^{(1)}_{\varkappa}\left(z_j|\bar z,\bar y\right)\prod_{j=1}^b\mathcal Y^{(2)}_{\varkappa}\left(y_j|\bar z,\bar y\right)}S,
\end{split}
\ee
where $S$ is defined in \eqref{S3}, $\mathcal N$ and $\mathcal M$ in \eqref{N}, \eqref{N_anti_diag} and \eqref{M}, \eqref{M_anti_diag}. We also used here that $\tau(c/2|\bar z;\bar y)$ is nothing but $\exp(ip(\bu))$ and $\tau_{\varkappa}(c/2|\bar z;\bar y)$ is $\varkappa_1\exp(ip(\bz))$. 

Note, that at $z\to\infty$, $y\to\infty$ integrand of \eqref{contour_corr_2} vanishes. Thus instead of evaluation of integrals  by the residues inside the integration contours we can evaluate integrals by the residue outside the integration contours. One can check that the only poles of the integrand  are the first order poles at $z_j=\uc_k$ or $z_j=\ub_k$, $j,k=1,\dots,a$ and $y_j=\vc_k$ or $y_j=\vb_k$, $j,k=1,\dots,b$. Then switching to integration over these poles and taking into account that in the fundamental model $r^{(i)}_2=1$, $r_1^{(1)}(\bv)=e^{imp(\bu)}$ we immediately arrive at \eqref{contour_corr}.

\section*{Conclusion}

The first result of the paper are new representations for the matrix elements of the generation functions of correlators \eqref{operator_Reshetikhin_gl3} and \eqref{operator_Reshetikhin_gl21}. These are direct analogues of the Reshetikhin formula \eqref{scalp} derived in \cite{Res86} for the scalar products. We restricted ourselves here only to cases of $R$-matrices with algebra symmetry $\mathfrak{gl}(3)$ or $\mathfrak{gl}(2|1)$, but we would argue that absolutely the same formula holds for $R$-matrices with the arbitrary $\mathfrak{gl}(m|n)$ algebra symmetry. This is easy to proof using the same simple steps as it was done in section \ref{Reshetikhin} starting from the analogues of \eqref{partial_model}, \eqref{dual_partial_model} and  \eqref{scalp} that are known for arbitrary $\mathfrak{gl}(m|n)$ algebra symmetry (see \cite{HyperReshetikhin}). We leave this exercise to the interested reader.

The second result is integral representation \eqref{contour_corr}. In the algebra symmetry $\mathfrak{gl}(2)$ related models such integral representation allowed to establish the asymptotic behaviour of correlation functions \cite{KitanineMaiilletTerrasSlavnov2008, Kozslowski3}. We have a hope that the new representation will be suitable for the computation of asymptotic in the algebra symmetry $\mathfrak{gl}(2|1)$ related model. We also have a hope that the similar representation can be established for the case of arbitrary $\mathfrak{gl}(m|n)$ algebra symmetry. At least it is clear that integral representation can be written for form factors of generator $Q$  in case of $\mathfrak{gl}(3)$ algebra symmetry.

Let us also mention also that in the algebra symmetry $\mathfrak{gl}(2)$ related model integral representation similar to \eqref{contour_corr} was written also for dynamical case\footnote{Being precise, this result was proven there only for the fundamental model but we expect that it is true for arbitrary models related to the algebra symmetry $\mathfrak{gl}(2)$.} \cite{Q_series_dynamical}. Without given too much details but rather relaying on complete analogy with algebra symmetry $\mathfrak{gl}(2)$ case we would argue that dynamical correlation function can be computed using analogue of \eqref{contour_corr} in algebra symmetry $\mathfrak{gl}(2|1)$ related models  too. The result is quite similar to the static case and the only modification consist in presence of the additional factor $e^{itE(\bz;\by)}$ (of course this is nothing but eigenvalue of time propagation operator) under the integral  and modification of of integration contours by adding poles of this factor $\{\mathfrak e,\mathfrak o\}$ inside the contours
\be{contour_corr_gl21dynamic}
\begin{split}
\langle \buc;\bvc |\exp(\alpha Q) |\bub;\bvb \rangle
=\frac{1}{a!b!}\oint\limits_{\buc\cup\bub\cup\mathfrak e}\prod_{j=1}^a\frac{dz_j}{2\pi i}&\oint\limits_{\bvc\cup\bvb\cup\mathfrak o}\prod_{j=1}^b\frac{d y_j}{2\pi i}e^{-it(E(\bz;\by)-E(\bu;\bv))}\\
\times \varkappa_1^a\varkappa_2^b\frac{r_1^{(1)}(\bz)r_3^{(2)}(\by)}{r_1^{(1)}(\bub)r_3^{(2)}(\bvb)}&\frac{\det_{a,b}\mathcal N\left(\bar z,\bub|\bvb,\bar y\right)\det_{a,b}\mathcal N\left(\buc,\bar z|\bar y,\bvc\right)}{\prod_{j=1}^a \mathcal Y^{(1)}_{\alpha}\left(z_j|\bar z,\bar y\right)\prod_{j=1}^b\mathcal Y^{(2)}_{\alpha}\left(y_j|\bar z,\bar y\right)}S.
\end{split}
\ee
All notations here are the same as in \eqref{contour_corr}.

\subsection*{Acknowledgments}

Authors are grateful to F.~G\"ohmann, N.~Slavnov and K.~Kozlowski for fruitful discussions. 

\providecommand{\bysame}{\leavevmode\hbox to3em{\hrulefill}\thinspace}

\end{document}